\documentclass{aa}
\usepackage{psfig,caption2}

\psfigurepath{/gagax6/ur4/siess/LATEX/ARTICLE/PMS}

\def\log{\mbox{log}}
\def\ln{\mbox{ln}}

\def\msun{$\,\mbox{M}_\odot$}
\def\Msun{\mbox{M}_\odot}
\def\lsun{L$_\odot$}
\def\rsun{\,R$_\odot$}
\def\teff{$T_{\mathrm{eff}}$}
\def\Teff{T_{\mathrm{eff}}}
\def\geff{$g_{\mathrm{eff}}\ $}
\def\Geff{g_{\mathrm{eff}}}
\def\lnf{ln$f$}

\newcommand {\chem}[1] {${}^{#1}$}
\newcommand {\mio}[3]{#1^{#2{\scriptstyle #3}}}
\newcommand {\io}[3]{#1$^{#2{\scriptstyle #3}}$}
\newcommand {\mab}[3]{N_{\scriptstyle \mio{#1}{#2}{#3}}}

\def\hm{H$^{-}$}
\def\hp{H$^{+}$}
\def\hd{H$_{2}$}

\begin{document}

\thesaurus{08(02.05.2; 08.05.3; 08.08.1 ; 08.16.5, 04.01.1)}

\title{An internet server for pre-main sequence tracks of low- and
  intermediate-mass stars} 
\titlerunning{A internet server for pre-main sequence tracks}

\author{Lionel Siess \and Emmanuel Dufour\and Manuel Forestini}

\offprints{Lionel Siess}

\institute{Observatoire de Grenoble, Laboratoire d'Astrophysique,
Universit\'e Joseph Fourier, BP 53, F-38041, Grenoble Cedex 9, France}

\date{Received date, accepted date}

\maketitle

\begin{abstract}
  
  We present new grids of pre-main sequence (PMS) tracks for stars in the
  mass range 0.1 to 7.0\msun. The computations were performed for four
  different metallicities ($Z$=0.01, 0.02, 0.03 and 0.04). A fifth table
  has been computed for the solar composition ($Z$=0.02), including a
  moderate overshooting. We describe the update in the physics of the
  Grenoble stellar evolution code which concerns mostly changes in the
  equation of state (EOS) adopting the formalism proposed by Pols et al.
  (1995) and in the treatment of the boundary condition.  Comparisons of
  our models with other grids demonstrate the validity of this EOS in the
  domain of very low-mass stars. Finally, we present a new server dedicated
  to PMS stellar evolution which allows the determination of stellar
  parameters from observational data, the calculation of isochrones, the
  retrieval of evolutionary files and the possibility to generate graphic
  outputs.
 
\end{abstract}

\section{Introduction}

Interests in modeling of very low-mass stars (VLMS) has grown steadily over
the past decade due to big efforts in both theory and observations. Brown
dwarfs and very low-mass stars are now frequently discovered and the
abundant literature in this domain attests of the vitality of this field.
From these observations, we now realize that very low-mass stars represent
a large fraction of the stellar population. However, the precise
determination of the initial mass function, and more generally of stellar
parameters and evolutionary states is still limited by the accuracy of
stellar models. It is therefore desirable that different tracks be
available in order to understand the theoretical discrepancies and also to
estimate the uncertainties associated with the derivation of stellar
parameters.

On the theoretical side, the computation of the structure of very low-mass
stars is very challenging since it involves the micro-physics associated
with dense, cold and partially degenerate matter. In the interior of VLMS,
collective effects due to large densities become important and an accurate
treatment of pressure ionization and Coulomb interactions is required.
Also, in the cool atmosphere of VLMS, several molecules form and produce
strong absorption bands that considerably modify the emergent spectrum. As a
consequence, the stellar surface cannot be considered as a black body and
stellar atmosphere models must also be used.

Since our last grids (Siess et al. 1997, hereafter SFD97), substantial
improvements have been made in the Grenoble stellar evolution code that led
us to the production of new grids of PMS tracks. The main modification to
the code concerns the equation of state (EOS) which treats more accurately
the behavior of cold and degenerate matter. In particular, we now take into
account the effects associated with the pressure ionization and treat more
carefully degeneracy conditions in stellar interior. We also updated the
opacity tables using the latest release of the OPAL group and modified our
surface boundary conditions using analytic fits to stellar atmosphere
models. The updates to the code are describe in Sect. \ref{code} followed
by a brief description of the grids. Then in Sect. \ref{comp} we compare
our models to grids recently published by other groups.  Finally, we
describe a new server dedicated to PMS evolution.

\section{The stellar evolution code}
\label{code}

A general description of the stellar evolution code has already been
presented in several papers (i.e. Forestini 1994, SFD97) and only
subsequent modifications are discussed. They include update in the opacity
tables, in the treatments of the atmosphere and the incorporation of a new
formalism for the equation of state.

\subsection{Opacity tables}

At high temperature (i.e. above 8000K), we use the OPAL radiative opacities
by Iglesias and Rogers (1996) which now include additional metal elements.
The tables are interpolated in temperature, density and H, He, C, O and
metal mass fraction using the subroutine provided by the OPAL group.
However, for the computation of very low-mass stars ($M\la 0.2$\msun), we
had to extrapolate the opacity tables to $\log R = \log(\rho/T^3_6) = 4.0$
  but this is of little effect since the internal region concerned with
  this extrapolation ($T \ga 10^5$ K) has an almost adiabatic behavior.
  Consequently, the temperature gradient is governed by the thermodynamics
  coming from the equation of state and is nearly independent of the
  opacity coefficient.  Conductive opacities are computed from a modified
version of the Iben (1975) fits to the Hubbard and Lampe (1969) tables for
non-relativistic electrons, from Itoh et al. (1983) and Mitake et al.
(1984) for relativistic electrons and from formulae of Itoh et al.  (1984)
as well as Raikh and Yakovlov (1982) for solid plasmas. Below 8000K, we use
the atomic and molecular opacities provided by Alexander and Fergusson
(1994) as in SFD97.

\subsection{Treatment of the atmosphere}  
\label{atmos}

The stellar structure is integrated from the center to a very low optical
depth ($\tau = 0.005$) in the atmosphere. In the regions where $\tau < 10$,
we constrain the atmospheric temperature profile (as well as the radiative
pressure and gradient) to correspond to those coming from realistic
atmosphere models determined by the integration of the radiative transfer
equation.
  More specifically, we have performed one analytic fit of
  $T(\tau)$, as a function of effective temperature \teff, surface gravity
  \geff and metallicity $Z$ which agrees with the atmosphere models to
  within 20\% or less. The parameters of this fit are constrained by
  various atmosphere models, namely (1) Plez (1992) for $2500<\Teff<4000$K,
  (2) Eriksson (1994, private communication, using a physics similar to
  Bell et al. 1976) up to 5500K and (3) Kurucz's models (1991) computed
  with the ATLAS 12 code above 5500K. Unfortunately, the lack of atmosphere
  models in some regions of the $Z-\Geff$ plane did not allow us to
  precisely determine the behavior of this fit at high gravity and/or non
  solar metallicities. This can potentially affect the modeling of very
  low-mass stars but considering the weak dependence of our fit on $Z$ and
  \geff compared to the one on \teff, we decided to use this unique fit in
  all our computations instead of using a grey atmosphere
  approximation.
This fit also provides a faster convergence in
the surface layers due to the smooth profile of $T(\tau)$ and of its
derivatives. Finally, let us mention that we define the effective
temperature as $T_{\mathrm{eff}} = (L/4\pi \sigma R^2)^{1/4}$ where $L$ and
$R$ are estimated at $\tau =2/3$.
  
\subsection{Equation of state}

In order to improve the description of the micro-physics relevant to VLMS,
we have revisited the computation of the equation of state (EOS), adopting
the scheme developed by Pols et al. (1995, hereafter PTEH). This new
formalism is characterized by three main points :
\begin{itemize}
\item we adopt the chemical point of view (i.e. Saumon et al. 1995) where
  bound configurations (ions, molecules) are populated according to
  matter-photons and matter-matter interactions,
\item we use the assumption of local thermodynamic equilibrium which allows
  the use of the Saha equations to determine the ion abundances as a
  function of the local values of temperature, $T$ and  number of electrons
  per unit mass, $N_e$, 
\item the plasma is described by three components, photons, ions and
    electrons, coupled only through the photo-ionization and
    photo-dissociation processes.  Each component of the plasma is
    represented by an independent term in the Helmoltz free energy
    (Fontaine et al. 1977). Non-ideal corrections due to Coulomb shielding
    and pressure ionization are treated separately in an additional term
    which expression has been fitted to the MHD EOS (Mihalas et al. 1988).
    This fit allows us to treat the regime of strong coupling while keeping
    the formalism of a separate additive term in the free energy to
    describe non-ideal effects (as explained in Saumon et al. 2000).
\end{itemize}

This new equation of state allows us to better take into account the effects
of pressure ionization, partially degenerate matter and Coulomb interactions
and provides very smooth profiles of the thermodynamic quantities since
analytical expressions of all their derivatives can be obtained.  Note that,
as we will see in the next section, the computation of this EOS naturally
introduces a new independent variable related to degeneracy.

\subsubsection{A new variable}

The general philosophy of this new EOS is to replace the density $\rho$ (or
pressure $P$) by a new independent variable which describes the electron
degeneracy.  To do so, we use the electron degeneracy parameter
$\eta_e=\beta\xi_e$ where $\beta=1/kT$ and $\xi_e$ is the electron chemical
potential. In this context, the density is given by
\begin{equation}
\label{defrho}
\rho=\frac{n_e(\eta_e,T)}{N_e(\eta_e,T,y_X)},
\end{equation}
where $n_e$, $N_e$ and $y_X$ are the number of electrons per unit volume,
per unit mass and the molar fraction of a chemical element $X$,
respectively.

$N_e$ depends on $\eta_e$ and $T$ through the generalized Saha equations,
and through the mass and charge conservation equations
\begin{eqnarray}
\label{sahae}
N_e &=  &\sum_{X,i} \,i\mab{X}{i}{+} \\
\label{saha}
\mab{X}{(i+1)}{+} &= 
 &\mab{X}{i}{+}\frac{{\cal{Z}}_{\mio{X}{(i+1)}{+}}}{{\cal{Z}}_{\mio{X}{i}{+}}} 
  e^{-\eta_e -\Delta \eta_e -\beta \chi^{i+}}\\
\label{sahah2}
N_{\mathrm H_{2}} &=
 &{\cal{Z}}_{\mathrm H_2} \biggl(\frac{N_{\mathrm H}}{{\cal{Z}}_{\mathrm H}}\biggr)^2\ \ \
 \mbox{for H}_{2}\\ 
 y_{X} {\cal{N}}_a &= & \sum_{i} \mab{X}{i}{+} \ \ \ \mbox{for
 X}\neq\mbox{H} \\ 
 y_{\,{\mathrm H}}\, {\cal{N}}_a &= & N_{\mathrm H} +  N_{\mathrm H^+} +2 N_{\mathrm H_{2}} +
 N_{\mathrm H^-}\ \ \ \mbox{for H}
\end{eqnarray}
where $\mab{X}{i}{+}$ represents the number per unit mass of the $i$ times
ionized atom \io{X}{i}{+} ($i$ can be negative for \hm\, for example),
$\chi^{i+}$ its ionization potential, ${\cal{Z}}_{\mio{X}{i}{+}}$ its
partition function and $A_X$ its atomic molar mass. $\Delta \eta_e$
  contains the contributions owing to the non-ideal effects of Coulomb
  interactions and pressure ionization which will be described in the
  following section.  The use of $\eta_e$ instead of $\rho$ as input
  variable for the Saha equations allows us to analytically derive the
  ionization states for all the desired elements anywhere in the star.

The number of electrons per unit volume $n_e$ is then obtained in terms of
$\eta_e$ and $T$ through the Fermi-Dirac integral
\begin{equation}
\label{FDI}
n_e = \frac{8\pi}{\lambda_{c}^3} \int_{0}^{\infty} 
          \frac{y^{2}dy}{e^{\frac{\displaystyle x}{T^*}-\eta_e}+1}\,
\end{equation}
where $y = \frac{p}{m_ec} = x^2+2x$, $\lambda_{c}$ is the electron Compton
wavelength and 
$T^*$=$kT/m_ec^{2}$.  This is a general
definition which allows for relativistic electrons. In order to speed up
the computation of $n_e$, we use 
polynomial fits to this integral. These fits have a simple form when
expressed in terms of two variables, called $f$ and $g$ by Eggleton et al.
(1973), arising from the asymptotic expansion of Eq. \ref{FDI}. Following
these authors,
\begin{equation}
\label{etaf}
\eta_e = \ln(f)+2\times(\sqrt{1+f}-\ln(1+\sqrt{1+f})).
\end{equation}
As a consequence, the variable ln($f$) (hereafter \lnf), very well suited
to describe the degeneracy of the plasma, is our new independent variable
replacing $\rho$ in the stellar structure equations. The change of variable
$\rho$\,(\lnf, $T$, $y_{i}$) is defined by Eqs. \ref{defrho}, \ref{sahae},
\ref{FDI} and \ref{etaf}. 
In weakly degenerate conditions, \lnf\ can be identified with a good 
approximation to the degeneracy parameter $\eta_e$ while at high degeneracy 
$f$ scales as $\eta_e$.
 
\subsubsection{Physical ingredients of the EOS}
\label{physeos}

The plasma is always considered as {\it partially} ionized for H, He, C, N,
O and Ne. The other species are considered to be either neutral or totally
ionized. Our tests indicate that for $M < 3$\msun, the structure and
evolution of the  stars are almost indistinguishable in both prescriptions
(as far as PMS is concerned). In the more massive stars ($M\ga 3$\msun) we
report that models assuming total ionization for the heavy elements have
initially a slightly shallower convective region but when the star reaches
the ZAMS, these differences vanish.

  Non-ideal effects include Coulomb shielding for all spe\-cies and
  pressure ionization for H, He, C, N and O, except for H$^{-}$. Both
  effects have been incorporated following PTEH by means of analytic fits
  to the non-ideal terms in the Helmoltz free energy. The fit to the
  pressure ionization has been derived from the MHD EOS for a mixture
  $X$=0.70 and $Z$=0.02. Nevertheless, they have been used in all our
  computations, regardless of the chemical composition. According to Pols
(private communication), this is a reasonable approximation since these
fits give the expected behavior of pressure ionization as a function of $T$
and of the degeneracy parameters ($\eta_e$ or \lnf).

\begin{figure}
\psfig{file=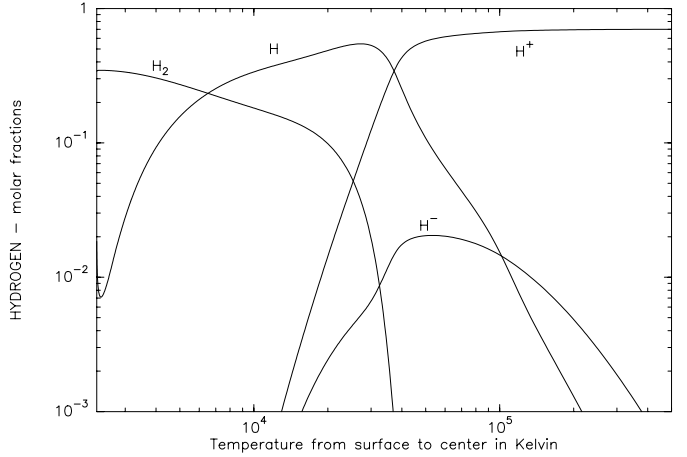,width=\columnwidth,angle=-90}
\caption{Profiles of H, \hp, \hm\, and \hd\ inside a 0.1\msun\ on the
  ZAMS. The star has an age of $8.36\times 10^9$ yr, a solar composition
  and is located in the HRD at $\Teff = 2776$\,K and $L = 8.2\times
  10^{-4}$\lsun}
\label{ioniz}
\end{figure}
The computation of \hd\ and \hm\, abundance is taken into account. For
  \hd, the partition function of Irwin (1987) is used in Eq. \ref{saha}
  below 16000K and had to be extrapolated at higher temperatures to ensure
  the convergence of the model. This is certainly of little importance
  since in this region \hd\, is insignificant. It must be noted that for
very low-mass stars ($M\la 0.4$\msun), \hd\ becomes very abundant in the
surface layers and must be included in the computations to ensure
consistency between the atmosphere models (which include this molecule) and
the EOS.  For example, in a 0.1\msun\ 50\% of the hydrogen is trapped into
\hd\ at a temperature of 11800K (Fig.  \ref{ioniz}). Further inside the
star, this molecule cannot survive.\\ Due to its negative charge, the
abundance of \hm\, derived from Eq. \ref{saha} is proportional to
$\exp(\eta_{e})$ instead of $\exp(-\eta_{e})$. This different
  dependence of \hm\, abundance on the degeneracy parameter is not
  reproduced by the fit to the pressure ionization, thus preventing for the
  moment the
  treatment of this effect on this ion.  This different behavior also
explains why the peak of \hm\, is located at the maximum of the degeneracy
parameter, around $T = 5\times 10^4$K where $\eta_{e}=3.02$.  We also
  verified that because of its small abundance in the encountered
  conditions, \hm\, has a negligible effect on the
  evolution tracks, even if its influence on the opacity coefficient can be
  large. Finally, in the central region of the star, H only subsists
in its ionization state H$^+$.

  Compared to Pols et al. (1995), our EOS presents slight modifications
  which mostly concerns the treatment of ionization. More specifically, now
  we compute the ionization states of C, N, O and Ne everywhere in the
  star, the elements for which ionization is not followed can be considered
  either completely ionized or neutral, the most recent partition function
  of Irwin (1987) is used for \hd\, instead of Vardya's (1960) and the
  presence of \hm\, has been accounted for in the Saha equations.  Apart
  from these modifications, the basic scheme and physical inputs (mainly
  the treatments of the non-ideal corrections and degeneracy) are the same.
  We checked that by adopting the same ionization treatment as PTEH
  (i.e. without the computation of \hm\, and assuming complete ionization
  for all the elements heavier than He), all the physical quantities
  derived form our EOS are comparable to the one computed by PTEH to within
  less than 1\% . Consequently, we refer the reader to these authors for a
  detailed comparison of this EOS with other ones.  In particular, they
  found that due to the use of a simplified treatment of pressure
  ionization, this EOS can differ by up to 10\% with the MHD and OPAL EOS,
  and even a bit more in the region of the $\rho - T$ plane delimited by
  the $3.0 < \log T < 4.5$ and $0 < \log \rho < 2$.

\section{Pre-Main sequence computations}
\label{grids}

\subsection{Initial models}

Our initial models are polytropic stars that have already converged once
with the code. The central temperature of all our initial models is below
$10^6$K so deuterium burning has not taken place yet. The stars are
completely convective except for $M \ge 6$\msun\ where a radiative core is
already present. For solar metallicity ($Z=0.02$), we use the Grevesse and
Noels (1993) metal distribution. For different $Z$, we scale the abundances
of the heavy elements in such a way that their relative abundances are the
same as in the solar mixture.

\subsection{The grids}

\begin{figure}
  \psfig{file=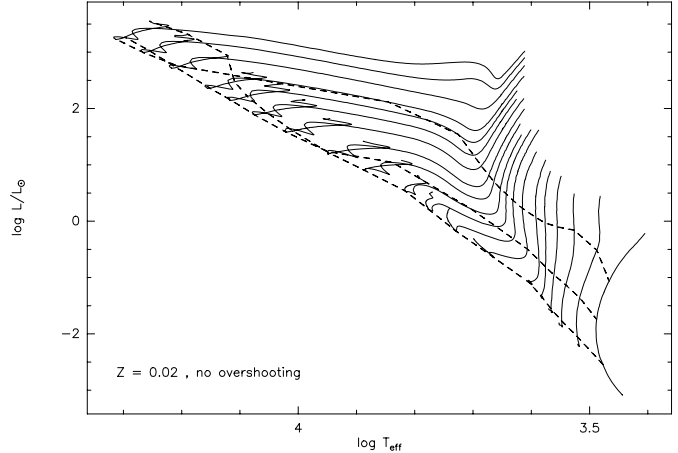,width=\columnwidth,angle=-90}
\caption[ ]{Evolutionary tracks from 7.0 to 0.1\msun\ for a solar
  metallicity ($Z$=0.02 and $Y$=0.28).  Isochrones corresponding to $10^6$,
  $10^7$ and $10^8$ (dashed lines) are also represented. This figure has
  been generated using our internet server}
\label{fighrd}
\end{figure}

Our extended grids of models includes 29 mass tracks spanning the mass
range 0.1 to 7.0\msun. Five grids were computed for four different
metallicities encompassing most of the observed galactic clusters
($Z=0.01$, 0.02, 0.03 and 0.04) and, for the solar
composition ($Z=0.02$), we also computed a grid with overshooting,
characterized by $d = 0.2H_p$, where $d$ represents the distance (in
unit of the pressure scale height $H_p$ measured at the boundary of the
convective region) over which the convective region is artificially
extended.
An illustration of these grids is presented
in Fig. \ref{fighrd}. Our computations are standard in the sense that they
include neither rotation nor accretion. The Schwarzschild criterion for
convection is used to delimit the convective boundaries and we assume
instantaneous mixing inside each convective zone at each iteration during
the convergence process. 

During the PMS phase, the completely convective star contracts along its
Hayashi track until it develops a radiative core and finally, at central 
temperatures of the order of $10^7$K, H burning ignites in its center.  The
destruction of light elements such as \chem{2}H, \chem{7}Li and \chem{9}Be
also occurs during this evolutionary phase. Deuterium is the first element
to be destroyed at temperatures of the order of $10^6$K. The nuclear energy
release through the reaction \chem{2}H(p,$\gamma$)\chem{3}He temporarily
slows down the contraction of the star. Then, \chem{7}Li is burnt at $\sim
3\times 10^6$K shortly followed by \chem{9}Be.  For a solar mixture, our models
indicate that stars with mass $\le 0.4$\msun\ remain completely convective
during all their evolution. We also report that in the absence of mixing
mechanisms, stars with $M > 1.1$\msun\ never burn more than 30\% of their
initial \chem{7}Li.

The account for a moderate overshooting characterized by $d = 0.20 H_p$
significantly increases the duration of the main sequence (MS) for stars
possessing a convective core and provides additional Li depletion during
the PMS phase.  More quantitatively, the MS lifetime of a 1.2\msun\ star is
increased by 25\%, this percentage then decreases to level off around a
15\% increase for $M \ga 3$\msun. Surface depletion of Li occurs only in
stars with $0.4 < M \la 1.3$ for which the temperature at the base of the
convective envelope can reach $3\times 10^6$K. Our models including
overshooting indicates that Li is much more efficiently depleted around
0.8\msun, where its abundance is $\sim 3\times 10^5$ smaller than in a
standard evolution.

\subsection{Our fitted solar model}

With the above physics, we have fitted the solar radius, luminosity and
effective temperature to better than 0.1\% with the MLT parameter $\alpha =
1.605$ and an initial composition $Y= 0.279$ and $Z/X = 0.0249$ that is 
compatible with observations. The values used for this fit are quite
similar to those obtained by other modern stellar evolution codes (see e.g.
Brun et al.  1998). Our sun, computed in the standard way, i.e. without the 
inclusion of any diffusion processes, has the following internal features 
\begin{itemize}
\item at the center : $T_c = 1.552\times 10^7$K, $\rho_c =
  145.7$\,g\,cm$^{-3}$, $Y_c = 0.6187$ and the degeneracy parameter $\eta_c
  = -1.527$;
\item at the base of the convective enveloppe : $M_{\mathrm{base}} =
  0.981999$\msun, $R_{\mathrm{base}} =0.73322$\rsun, $T_{\mathrm{base}} =
  1.999\times 10^6$K and $\rho_{\mathrm{base}} = 1.396 10^{-2}$\,g\,cm$^{-3}$.
\end{itemize}
These numbers are in very close agreement with other recently published
standard models (see e.g. Brun et al. 1998, Morel et al. 1999, Bahcall \&
Pinsonneault 1996). 

  In our grid computed with overshooting, we found that the 1 \msun
  model maintains a small convective core of $\sim 0.05$\msun during the
  central H burning phase. This would consequently be the case for the
  fitted sun but present day helioseismic observations cannot exclude this
  possibility (e.g. Provost et al.  2000).

\section{Comparison with other works}
\label{comp}

In this section we compare our PMS tracks with the computations made
available by the groups listed in Table \ref{tab1}. These comparisons show
the accuracy of the EOS and the pertinence of our models in the domain of VLMS.
\begin{table}
\caption[]{PMS stellar models used in the comparisons}
\begin{flushleft}
\begin{tabular}{lllll}
\noalign{\smallskip}\hline\noalign{\smallskip}
 X &  Z &   group & code\\
0.725 & 0.019 & Baraffe et al. (1998) & BCAH\\
0.680 & 0.020 & Charbonnel et al. (1999) & Geneva\\
0.691 & 0.019 & D'Antona \& Mazzitelli (1997) & DM97\\
0.699 & 0.019 & Swenson et al. (1994) & SW94\\
0.703 & 0.020 & Siess et al. (1997) & SFD97\\
0.703 & 0.020 & this work \\
\noalign{\smallskip}
\hline
\end{tabular}
\end{flushleft}
\label{tab1}
\end{table}

Pre-main sequence tracks differ from one group to another due to
differences in the constitutive physics (EOS, convection), but also in the
treatment of the surface boundary conditions. In the last decade, a
tremendous amount of work has been done to better understand the physical
processes taking place in the extreme regime of high density and low
temperature encountered in the interior of VLMS. Three main EOS have
emerged: the MHD EOS used by the Geneva group and by DM97, the SCVH EOS
(Saumon et al.  1995, hereafter SCVH95) used by Baraffe et al. (1998,
hereafter BCAH) and the OPAL EOS (Rogers et al.  1996) used by DM97 in the
regime of high temperature. Comparisons between these different EOS
(SCVH95, Rogers et al. 1996, Chabrier \& Baraffe 1997, Charbonnel et al.
1999) showed a good agreement within their domain of validity. For
example the MHD and SCVH EOS give very similar results in the mass range
$0.4 \la M \la 0.8$\msun.

\begin{figure}
\psfig{file=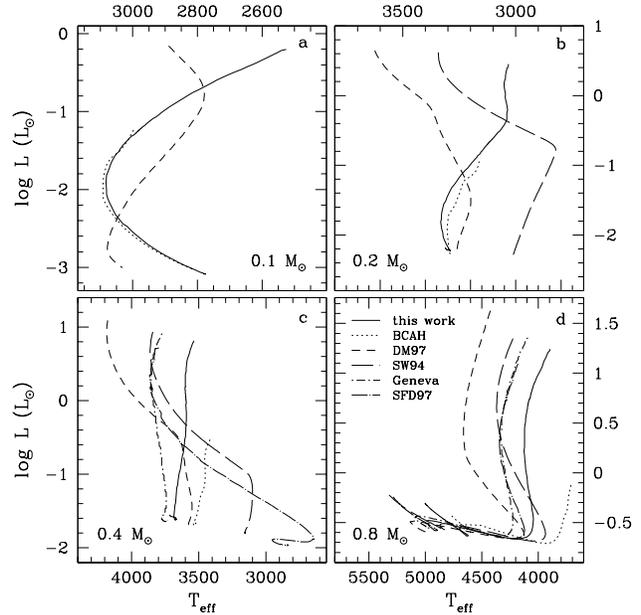,width=\columnwidth}
\caption[ ]{HR diagrams comparing the different PMS tracks for the models
  listed in Table \ref{tab1}}
\label{figcomp}
\end{figure}

Figure \ref{figcomp} shows that, the morphology of our tracks is very
similar to the BCAH and Geneva models, for all the considered masses.  This
is a strong evidence that we correctly follow the thermodynamics involved
in these objects. This is not surprising since, as mentioned in Sect.
\ref{physeos}, we treat the non-ideal effects through analytic fits to the
MHD EOS. In comparison, our previous tracks (long dashed-dotted lines) were
systematically too red and their paths were very different from the new
ones. This is due in part to the fact that \hd\ was not accounted for in the
computations and also because the treatment of the non-ideal effects was
less accurate. As a consequence, the internal structure of our old models
was less centrally condensed and the stellar radius slightly larger
compared to our new models.

At the lower mass end ($M$=0.10 and 0.20\msun), the similarity with BCAH
models is striking, as effective temperature differences are $\la 50$K
(much less than observational uncertainties along the Hayashi tracks!). In
particular, this shows that our boundary conditions (Sect.  \ref{atmos})
are in reasonnably good agreement with the much more sophisticated
atmosphere models used by BCAH.
We see however that in the domain of very low mass stars ($M \la 0.4$\msun)
strong morphological differences are present. More specifically, we observe
(Fig.  \ref{figcomp}) that the DM97 and SW94 tracks present a strong
inflexion in their Hayashi path which is not reproduced by the other
groups. Several reasons can be put forward to explain these differences.
First DM97 uses a convection model which differs substantially from the
commonly used standard MLT prescription. As shown by D'antona and
Mazzitelli (1998), this can result in \teff\ differences as large as $\sim
200$K along the Hayashi line (see also below).  Secondly, in this mass
range the morphology of the evolutionary tracks is particularly sensitive
to the EOS and to the different treatments of the thermodynamics. Finally,
their use of a grey atmosphere approximation is certainly not satisfactory
for modeling cool stars whose emergent spectrum diverges significantly from
a black body.
In order to estimate the impact of using a grey atmosphere treatment in our
stellar evolution code, we have also computed some additional tracks for
different masses. We report that the non-grey atmosphere models are
systematically cooler than the grey ones, as a result of molecular
blanketing in the outermost layers (e.g. Chabrier et al. 1996). This
probably also explains why the Geneva models, which use the same treatment
of convection with a similar value for MLT parameter ($\alpha= 1.6 \pm
0.1$), are systematically hotter than ours.

For higher mass stars ($M \ga 0.6$\msun), the morphology of the tracks
becomes quite similar as the non-ideal physical effects and the molecular
absorption in the atmosphere vanish. The shifts in \teff\ between the
different tracks mainly result from the different chemical compositions
and/or values of $\alpha$ in the MLT and can be as large as 200K. In
particular, the Hayashi tracks of the BCAH models become increasingly
cooler as the stellar mass increases compared to all the other tracks. This
most probably results from their smaller value of the MLT parameter
($\alpha= 1$), the effect of which is only significant above $\sim
0.6\Msun$ (Chabrier and Baraffe 1997) and is to shift the tracks towards
lower \teff. The fact that species heavier than H and He are not accounted
for in the BCAH EOS could be part of the explanation as well.  Finally, for
the most massive stars ($M> 1.0$\msun), the morphology of the tracks is
very similar. We notice however that our models are slightly cooler than
the DM97 tracks by $\sim 100$K.

A comparison of the isochrones shows that our tracks are generally slightly
more luminous than those of others. Consequently, for a given luminosity,
the stellar age estimated from our models will be smaller, the relative
effects being more pronounced for younger stars. The discrepancy between
the different isochrones is particularly strong below $10^6$ yr, where an
age determination remains, in any case, very misleading. This is in part
due to the fact that early in the evolution, the results still depends on
the initial state of the star (Tout et al. 2000).

Aside from the above-mentioned explanations for the discrepancies between
the tracks, other physical ingredients usually not reported in the
literature may also affect the evolutionary path of a PMS stars.
Consequently, the origin of some of the observed discrepancies cannot be
clearly identified. To illustrate this purpose, we will now demonstrate
that, even within the MLT, different prescriptions to solve the third
polymonial equation for the convective gradient give rise to different
degrees of superadiabadicity and thus different location of the star in the
HRD. Indeed, the \teff\ location of a fully convective PMS stars is very
sensitive to the degree of superadiabadicity in the sub-photospheric layers
which in turn depends on the prescription used to compute the convective
temperature gradient (see e.g. D'antona \& Mazzitelli 1998). We performed a
series of tests using two different formalisms for the computation of the
convective gradient, namely the Cox (1984, Chap.  14) and the Kippenhahn
and Weigert (1991, Chap. 7) formalisms. The latter, used in our grids,
leads in the superadiabatic region to a temperature gradient
$\sim 20\%$ higher than the one 
coming out 
from the Cox formalism.  This translates into effective temperature shifts
of the Hayashi tracks of 100 to 300K, depending on the stellar mass.  The
Cox solutions have systematically smaller radii and the Hayashi lines are
consequently bluer. Finally with the receding of the convective envelope,
the degree of superadiabadicity decreases and the temperature shift reduces
to $\sim 100$K. These differences mainly result from the assumed ``form
  factor'' used in the MLT formalism. Indeed, the shape of the convective cells
  directly affects the radiative loss efficiency and a change in
  the form factor can be compensated by a modification to the $\alpha$
  parameter. We refer the reader to Henyey et al. (1965) for a detailed
  discussion of the arbitrariness of some constants used in the MLT.
   Among other numerical tests, let us also mention that
the evolution of the structure of PMS stars does not depend at all on the
prescription used to write the gravothermal energy production rate
$\varepsilon_{\mathrm grav}$ (i.e.  using the internal energy, the entropy
or the pressure; see Kippenhahn \& Weigert 1991, Chap. 4). This is due
  to  our completely consistent treatment of the thermodynamics with our
  EOS formalism.

Given the relatively good agreement between the different EOS involved in
this comparison, it comes out that the physical and numerical treatment of
convection is certainly the most influent parameter for the modeling of
fully convective PMS. It can account for \teff\ differences as large as
300K. Faced to our poor knowledge of convection, large discrepancies
between the different sets of tracks are unfortunately inevitable.

\section{Internet interface to stellar model requests}

The internet site located at {\sf
  http://www-laog.obs.ujf\--greno\-ble.fr/\-activites/\-starevol/evol.html}
offers several services to take advantage of our stellar model database.
Among these facilities, it is possible to compute an isochrone, to
determine the stellar parameters of a star knowing its position in the HR
diagram, to draw a specific HR diagram and of course, to retrieve all the
evolutionary files.

For the computation of isochrones, the user specifies the metallicity, ages
and selects the mass tracks entering the computation of the isochrone.  The
results are displayed on the screen and can be saved in a file.  We also
give the luminosity, effective temperature, radius as well as the colors
and magnitudes corresponding to the location of a given star at any
specified age. We use the conversion table provided by Kenyon and Hartmann
(1995) and display the different colors in the Cousin system.  It is also
possible to ask for further information and get, for example, the surface Li
abundance, the central temperature or the extent of the convective envelope
for the set of stars selected to compute the isochrones.

The other facility is the determination of stellar parameters. The user
specifies the coordinates of a star in the HR diagram using either the
luminosity or magnitude and \teff\ of colors and then, the program computes
the theoretical track that passes through this observational point.  This
procedure allows the determination of the stellar age, mass and radius
assuming the star is on its pre-main sequence track. If requested,
additional information such as surface chemical composition, central
properties or moments of inertia can also be provided.

The third facility deals with graphic outputs. In this page, the user can
plot HR diagrams in any combination of color and magnitude. It is also 
possible to zoom in by stating limits, to superpose isochrones
and/or the ZAMS.

\section{Conclusion}

We presented new pre-main sequence evolutionary tracks for low- and
intermediate-mass stars.  Comparisons of our models with other PMS tracks
indicate rather strong discrepancies in the regime of very low-mass stars
($M\la 0.5$\msun). Our tracks are very similar, in morphology and effective
temperature, to those computed by Baraffe et al. (1998) and the Geneva group
(Charbonnel et al, 1999). This similarity is a strong indicator that our
treatment of the EOS and boundary conditions are correct. Conversely, we
report strong deviations of these tracks with the D'Antona and Mazzitelli
(1997) ones, especially below 0.3 \msun. For higher mass stars, the
morphology of the tracks are similar and effective temperature differences
$\la 200$K are noted, partly due to different mixing length parameter values
and chemical compositions.  Comparisons of the isochrones indicate age
determination remains, in any case, very uncertain below $10^6$yr.
Finally, we present our 
internet server which provides several facilities to use and take advantage
of our large database of PMS stellar models.

\begin{acknowledgement}
  The authors wish to thank C. Tout and O. Pols for their numerous and very
  helpful interactions during the implementation of the EOS in the code. LS
  also wants to thank F. Roch for her help in building the server. 
  The computations presented in this paper were performed at the ``Centre
  de Calcul de l'Observatoire de Grenoble'' and at ``IMAG'' on a IBM SP1
  computer financed by the MESR, CNRS and R\'egion Rh\^one-Alpes.
\end{acknowledgement}

\end{document}